\documentclass[aps,prl,twocolumn,groupedaddress]{revtex4-1}

\usepackage{graphicx}
\usepackage{amsmath}
\usepackage{subfigure}

\begin{document}

\title{Optical pulse-shaping for internal cooling of molecules}
\author{C-Y. Lien}
\author{S. R. Williams}
\author{B. Odom}
\affiliation{Department of Physics and Astronomy, Northwestern University
\\ 2145 Sheridan Rd., Evanston, IL 60208}

\begin{abstract}
We consider the use of pulse-shaped broadband femtosecond lasers to optically cool rotational and vibrational degrees of freedom of molecules. Since this approach relies on cooling rotational and vibrational quanta by exciting an electronic transition, it is most easily applicable to molecules with similar ground and excited potential energy surfaces, such that the vibrational state is usually unchanged during electronic relaxation. Compared with schemes that cool rotations by exciting vibrations, this approach achieves internal cooling on the orders-of-magnitude faster electronic decay timescale and is potentially applicable to apolar molecules. For AlH$^{+}$, a candidate species, a rate-equation simulation indicates that rovibrational equilibrium should be achievable in 8 $\mu$s. In addition, we report laboratory demonstration of optical pulse shaping with sufficient resolution and power for rotational cooling of AlH$^+$.
\end{abstract}

\maketitle

\section*{Introduction}
Optically controlling and manipulating the external and internal degrees of freedom of molecules has aroused wide interest in the physics and chemistry communities\cite{Vogelius2004, Viteau2008, Hudson2009, Herschbach2009, Shuman2010, Vogelius2006, Morigi2007, Kowalewski2007, Zeppenfeld2009}. The additional rotational and vibrational internal structure of molecules as compared with atoms complicates laser cooling schemes. For typical diatomic molecules at room temperature, the population is distributed among many rotational and vibrational levels, each requiring an optical pumping laser at a unique wavelength in order to prepare the sample in the rovibrational ground state.

There are currently two experimentally demonstrated approaches from the groups of M. Drewsen and S. Schiller to optically cool trapped polar molecular ions into the rovibrational ground state\cite{Staanum2010, Schneider2010}. Both drive specific vibrational transitions while relying on blackbody radiation (BBR) to redistribute the population, achieving fractional rovibrational ground state populations about an order of magnitude higher than the room temperature equilibrium distribution. However, this vibrational excitation approach is much less efficient for molecules of larger reduced mass $\mu$ because the vibrational relaxation time scales roughly as $\mu^{2}$. Furthermore, the BBR redistribution method is only efficient for hydride molecules because their rotational spacings are on order of 20 cm$^{-1}$ ($k_{B} \times$30 K), which is sufficiently large to still be driven on seconds timescales by the low-energy tail of the 300 K BBR spectrum\cite{Tong2011}. For non-hydride molecules such as ClF$^+$, BrCl$^+$, and SiO$^{+}$, the spacings of rotational states are much smaller, and the BBR redistribution timescales are on order of tens of seconds\cite{Tong2011, Nguyen2011}. 

An alternative method demonstrated by the group of S. Willitsch loads molecular ions (N$_2^+$) directly into a trap in specific rovibrational states\cite{Tong2010}. Several groups have investigated coherent population transfer of loosely bound Feshbach states into the rovibrational ground state\cite{Ni2008, Shi2010}; however, since there is no dissipation ($e.g.$, spontaneous emission) in these schemes, state transfer does not cool the internal degrees of freedom of molecules from a thermal distribution. 

Optical cooling by a pulse-shaped femtosecond laser\cite{Weiner2000} (PFL) has been employed to cool the vibrational population of neutral alkali dimers created from ultracold atoms by the group of P. Pillet\cite{Viteau2008, Sofikitis2009}. The same group discusses the possibility of using optical pulse-shaping for rotational cooling of Cs$_{2}$\cite{Viteau2008, Sofikitis2009}, but the required resolution, because of the small rotational constant, is far higher than the 1 cm$^{-1}$ resolution they achieved.  In this article, we consider the prospects of using optical pulse shaping to rotationally cool molecules with relatively large rotational constants.  We require cooling on an electronic transition with fairly diagonal Franck-Condon factors (FCFs), such that electronic relaxation usually does not change the vibrational state. In particular, we analyze optical pulse-shaped rotational cooling of AlH$^{+}$ molecular ions held in a radiofrequency Paul trap. Our technique is also readily applicable to other species with diagonal FCFs, such as BH$^{+}$\cite{Nguyen2011a} and SiO$^{+}$\cite{Nguyen2011}.

\section*{Rotational cooling scheme}

The previously-demonstrated rotational cooling schemes\cite{Staanum2010, Schneider2010} used infrared CW lasers to drive a small number of vibrational transitions, requiring on the order of tens of seconds for polar hydride molecules in a 300 K environment to reach equilibrium. For hydrides, these schemes are limited to seconds timescales by BBR-induced rotational transition rates and are in principle limited by the vibrational relaxation times of tens of milliseconds (for multi-wavelength extensions not relying on BBR redistribution). Electronic transitions are generally much faster than rotational and vibrational transitions; for example, the decay A$^{2}\Pi (v'=0)$ $\rightarrow$ X$^{2}\Sigma^{+} (v''=0)$ for AlH$^{+}$ occurs with a time constant of 10$^{-7}$ s, as compared with the 10$^{-2}$ s for X$^{2}\Sigma^{+} (v''=1)$ $\rightarrow$ X$^{2}\Sigma^{+} (v''=0)$. Unlike for vibrational excitation approaches, extension of our scheme to molecules with larger reduced mass (non-hydrides, $e.g.$, SiO$^{+}$\cite{Nguyen2011}) is also possible without incurring longer cooling times.

\begin{figure}[ht]
\begin{center}
\mbox{
\subfigure[]{\includegraphics[height= 5.4cm]{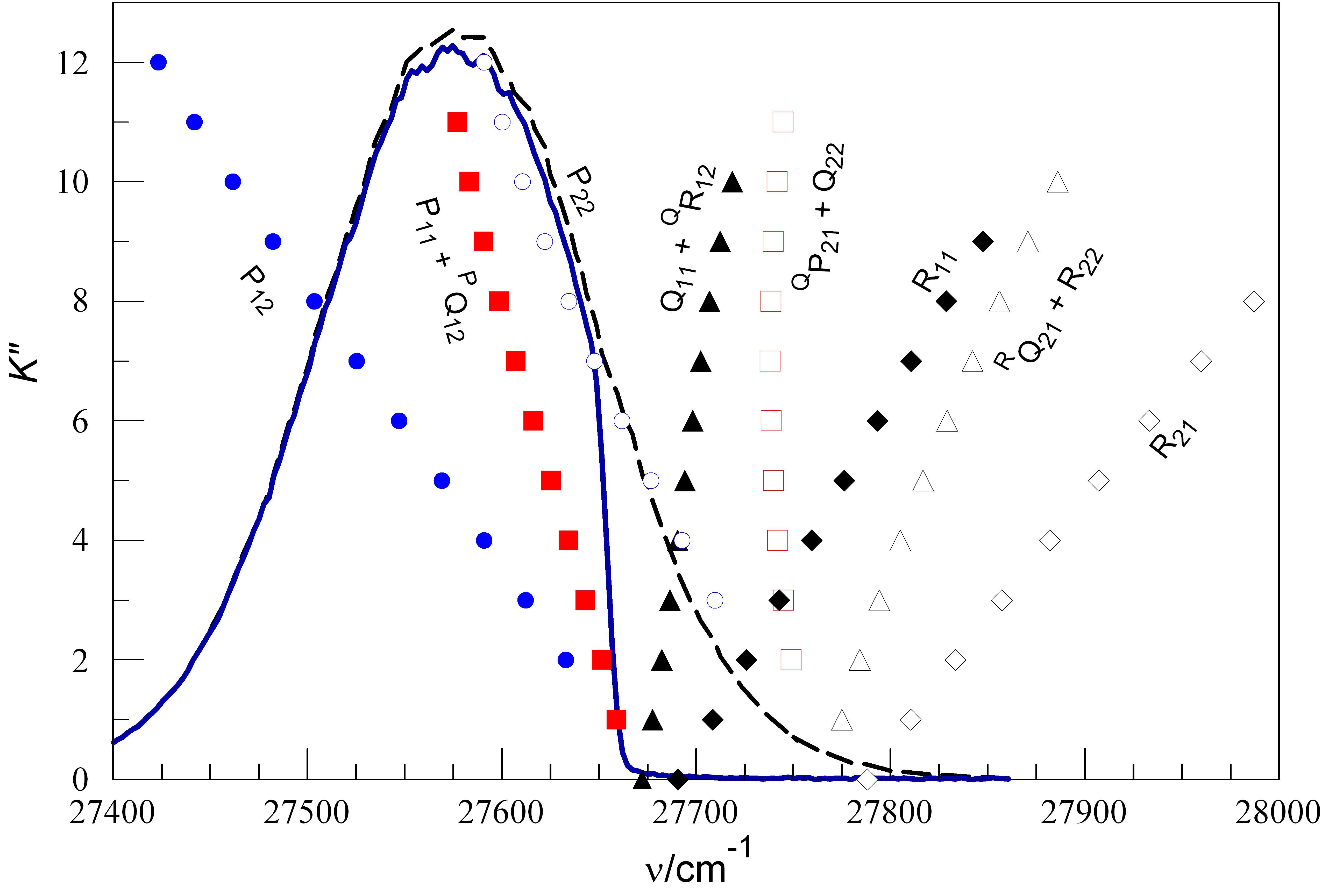}}
}
\mbox{
\subfigure[]{\includegraphics[height= 5.4cm]{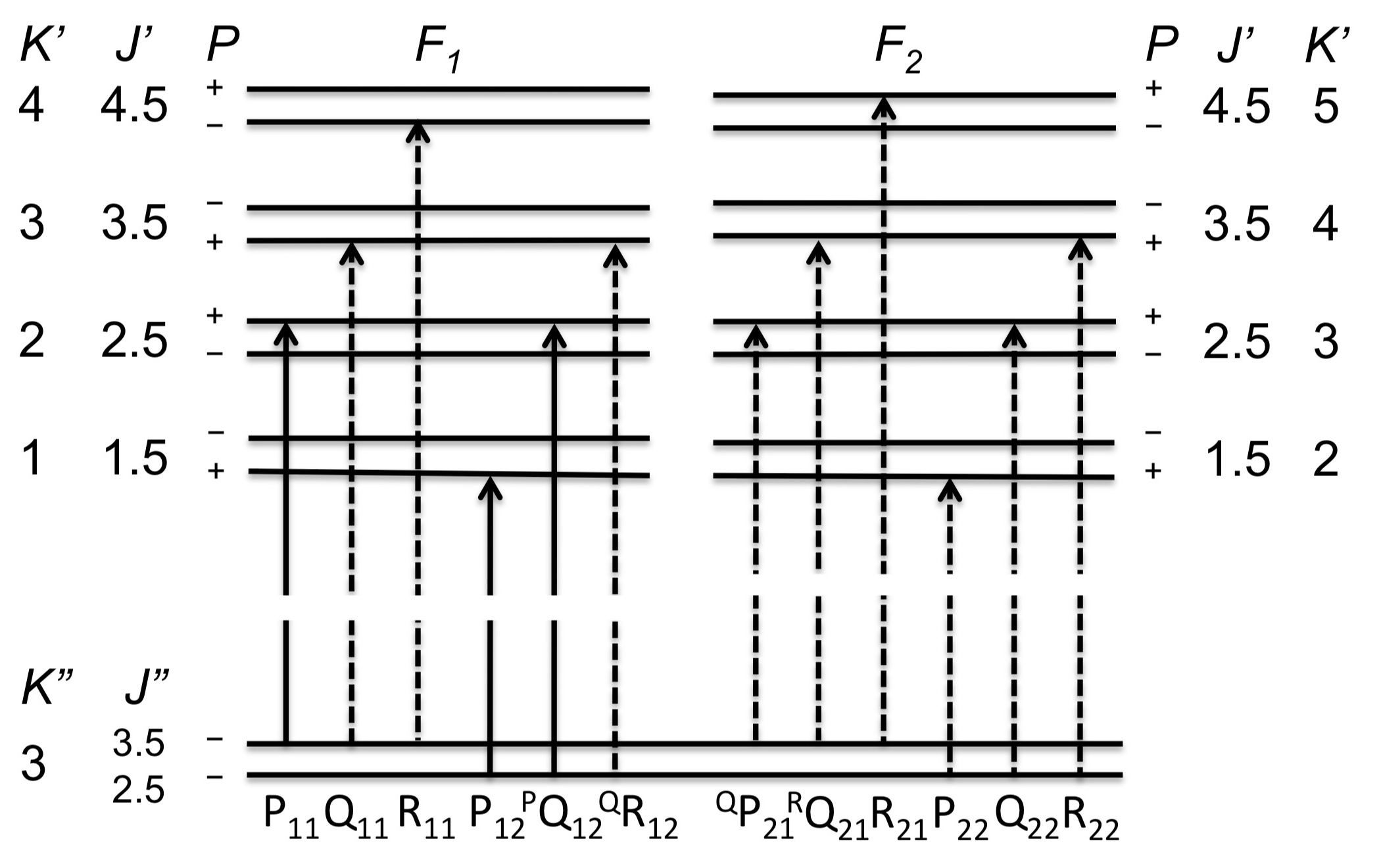}}
 }
\caption{PROC for AlH$^{+}$. (a) The pulse-shaped frequency-doubled femtosecond laser (solid-blue line, data) with a cutoff at 27,655 cm$^{-1}$ can be used for branch-selective optical pumping, cooling the rotational degree of freedom. The closed and open symbols represents the F$_{1}$ and F$_{2}$ transitions, respectively. The dashed line shows the femtosecond laser spectrum without pulse-shaping. The population among $K''=2-12$ rotational levels will be pumped to the dark states, $K''=0, 1$ (see text for details). (b) All transitions which could be driven from an example state, $K''=3$. Pulse shaping removes wavelengths corresponding to dashed lines in the figure. }
\end{center}
\end{figure} 

The pulsed-shaped rovibrational optical cooling (PROC) scheme proposed in this article uses a PFL to pump the population of diatomic molecular ions ($e.g.$, AlH$^+$) trapped in a linear Paul trap, to their rovibrational ground state. Because a PFL can pump many of the internal cooling transitions simultaneously, no BBR is required for redistributing the population among rotational states. Cooling is accomplished by a sequence of photon absorption events, each followed by spontaneous emission; throughout the process, the PFL spectrum is not adjusted. Since the highly diagonal FCFs of AlH$^+$\cite{Nguyen2011a} (see Table 1) limit the population leakage during electronic decay into higher vibrational states, only one PFL is required. For other molecules with less diagonal FCFs, a second PFL tuned to repump the excited vibrational populations could be added.

\begin{table}[h]
\small
  \caption{\ FCFs between A$^{2}\Pi (v'=0)$ and X$^{2}\Sigma^{+} (v''=0-2)$. The FCFs here are calculated using the LEVEL program\cite{LEVEL} (version 8.0) with the potential energy curves and dipole moments from Klein $et~al.$\cite{Klein1982}.}
  \label{tbl:FCFs}
  \begin{tabular*}{0.5\textwidth}{@{\extracolsep{\fill}}llll}
    \hline
      X$^{2}\Sigma^{+}$ & $v''=0$ & $v''=1$ & $v''=2$ \\
    \hline
     FCF & 0.99934 & 0.00060 & 0.00006 \\
    \hline
  \end{tabular*}
\end{table}

In our laboratory, AlH$^+$ ions are trapped and sympathetically cooled to sub-Kelvin translational temperatures via their Coulomb interaction with laser-cooled co-trapped barium ions. Note that sympathetic cooling in ion traps achieves translational but not rovibrational molecular ion cooling\cite{Molhave2000, Roth2005}. Translational cooling is not generally necessary for PROC, and the approach could also be applied to neutral samples in molecular beams. The AlH$^{+}$ molecular ions can remain trapped for hundreds of minutes in a UHV system, providing ample time to reach equilibrium with the 300 K environment even after a hot loading process. At 300 K, the population of AlH$^{+}$ is almost completely in its vibrational ground state X$^{2}\Sigma^{+} (v''=0)$, and $>$99$\%$ is distributed among the $K''=0$ to $K''=12$ rotational states.

The AlH$^{+}$ A$^{2}\Pi (v'=0)$ $\rightarrow$ X$^{2}\Sigma^{+} (v''=0)$ spectra in Figure 1 are calculated using the molecular Hamiltonian provided in Almy and Watson's work\cite{Almy1934}, and the molecular constants calculated by Guest and Hlrst\cite{Guest1981} and Szajna and Zachwieja\cite{Szajna2011}. The selection rules between A$^{2}\Pi (v'=0)$ and X$^{2}\Sigma^{+} (v''=0)$ are $\Delta J=0, \pm 1$ and parity $+ \leftrightarrow -$. The PFL drives only selected P- and Q-branch transitions as shown in Figure 1, which will result in rotational cooling and not heating. This arrangement creates dark states: the two rotational ground states, X$^{2}\Sigma^{+} (v''=0,~K''=0)$ and X$^{2}\Sigma^{+} (v''=0,~K''=1)$. The population of the two dark states cannot be transferred from one to the other because of parity; however, the parity barrier could be surmounted by applying sub-millimeter (far IR) radiation tuned to the X$^{2}\Sigma^{+} (v''=0,~K''=2)$ $\leftrightarrow$ X$^{2}\Sigma^{+} (v''=0,~K''=1)$ transition.

A home-built pulse-shaping device, consisting of diffraction gratings and cylindrical lenses in the 4-f Fourier-transform optical layout and a razor blade mask, is used to achieve amplitude tuning\cite{Weiner2000, Sofikitis2009a}. Figure 1 demonstrates our current pulse-shaping resolution to be better than 10 cm$^{-1}$, where the measurement is limited by the resolution of our spectrometer (Ocean Optics HR4000).

On the Fourier plane, the spot size of any individual spectral color incident on the mask is given by\cite{Weiner2000}
\begin{equation}
w_0=\frac{\cos{\theta_{in}}}{\cos{\theta_{d}}}\frac{\lambda f}{\pi w_{in}},
\end{equation}
and the spatial dispersion (with units of distance/frequency) is given by\cite{Weiner2000}
\begin{equation}
\alpha = \frac{\lambda^2 f}{c d \cos{\theta_d}},
\end{equation}
where $\lambda$ is the wavelength, $\theta_{in}$ and $\theta_d$ are the incident and reflected angles from the grating, $f$ is the focal length of the lens, $d$ is the grating period, $c$ is the speed of light, and $w_{in}$ is the radius of the collimated beam incident on the grating. 

The diffraction-limited full width at half maximum (FWHM) spectral resolution of pulse-shaping is then given by\cite{Weiner2000}
\begin{equation}
\delta v_\text{d}=(\ln  2)^{1/2}\frac{w_0}{\alpha}=(\ln  2)^{1/2}\frac{\cos{\theta_{in}}}{\pi w_{in}}\frac{c d}{\lambda}.
\end{equation}
Eq. (3) shows that increasing $w_{in}$ and decreasing $d$ leads to better resolution, but this resolution is limited by the size of the optics. For a near-Littrow configuration ($\theta_{in}\sim\theta_{d}$) with $\lambda=360$ nm, $d=1/3600$ mm, and $w_{in}=25$ mm, $w_0$ and $\delta v_\text{d}$ are 4.6 $\mu$m and 0.08 cm$^{-1}$, respectively. 

The practical spectral resolution could be worse than $\delta v_\text{d}$ if the scale of mask features is limited to some size $x$. In that case, the mask-limited pulse-shaping resolution is given by
\begin{equation}
\delta v_\text{m}=(\ln  2)^{1/2}\frac{x}{\alpha}=(\ln  2)^{1/2}\frac{x c d}{\lambda^2 f} \cos{\theta_d}.
\end{equation}
This expression for $\delta v_\text{m}$ is valid if $x>w_0$. Otherwise, different spectral colors on the Fourier plane would overlap, resulting in a smearing of the shaped profile. If a mask with 30 $\mu m$ structure is used, $f=1~m$, and the parameters given above, the expected pulse-shaping resolution is 0.48 cm$^{-1}$, which is much better than the 10 cm$^{-1}$ resolution required for AlH$^{+}$ PROC. For rotational cooling of SiO$^{+}$ molecules, 3 cm$^{-1}$ resolution is required to pump only the cooling transitions (P-branch transitions in this case) while not exciting the heating transitions\cite{Nguyen2011, Mogi2002}. For  AlH$^+$, BH$^+$, and SiO$^+$, the cooling-transition branches are well separated from the heating-transition branches, which simplifies the pulse-shaping requirements to removing the upper or lower portion of the spectrum. For more complicated spectra, where heating and cooling branches partially overlap, a more complex mask is required.

\begin{figure}[ht]
\centering
\includegraphics[height= 5.5cm]{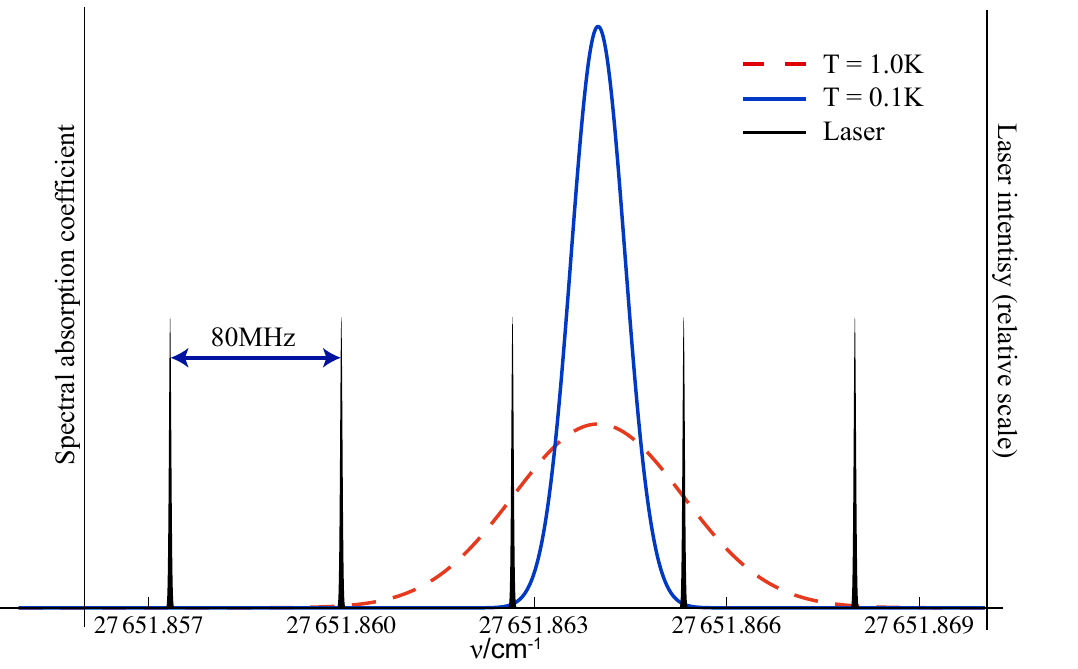}
\caption{Doppler broadening helps the overlap between the PFL and the transition linewidth}
\end{figure}

Our femtosecond laser (SpectraPhysics MaiTai) has an 80 MHz repetition rate, so each pulse has a comb-like frequency spectrum with frequency interval equal to the repetition rate. The linewidth of AlH$^{+}$  A$^{2}\Pi (v'=0)$ $\rightarrow$ X$^{2}\Sigma^{+} (v"=0)$ transitions is of order 2$\pi$ $\times$ 1/60 ns$^{-1}$ (2$\pi$ $\times$20 MHz), which is smaller than the comb interval; therefore, it would be possible for the PFL to miss some of the transitions we desire to pump. The simplest solution to this issue is to increase the translational temperature of the AlH$^+$/Ba$^+$ ion cloud (see Fig. 2), since a modestly Doppler-broadened linewidth ensures that the PFL can drive all the desired cooling transitions. One broadening approach is to directly excite the secular motion of the AlH$^{+}$ with an oscillating electric field. After internal cooling is finished, the secular drive is turned off, and the translationally hot molecular ions are again sympathetically cooled by the laser-cooled Ba$^+$. Another approach to temporarily increase the linewidth of the molecular ions is to push the entire ion cloud away from the geometric center of the Paul trap by applying a DC field. This displacement introduces excess micromotion\cite{Berkeland1998}, where the ions undergo swift oscillations at the RF frequency with amplitude proportional to the applied DC field. To reach the desired linewidth broadening (80 MHz) in the geometry of our trap ($r_0=3$ mm, $\Omega=2\pi\times3$ MHz), the ions have to be pushed only 2 $\mu$m away from the geometric trap center.

The typical optical transition linewidth for AlH$^+$ is 20 MHz with a saturation intensity of 70 $\mu$W~mm$^{-2}$. The frequency-doubled femtosecond laser has a power of 500 mW centered at 360 nm, with a 7 nm bandwidth. If the power were distributed over the 7 nm bandwidth evenly, the power available in the 20 MHz linewidth would be 3 $\mu$W. By focusing the laser onto a 100 $\mu$m diameter spot, a laser intensity of 300 $\mu$W~mm$^{-2}$ is achieved, which is a factor of 4 larger than the saturation intensity.

\begin{figure}[ht]
\begin{center}
\mbox{
\subfigure[]{\includegraphics[height= 5.4cm]{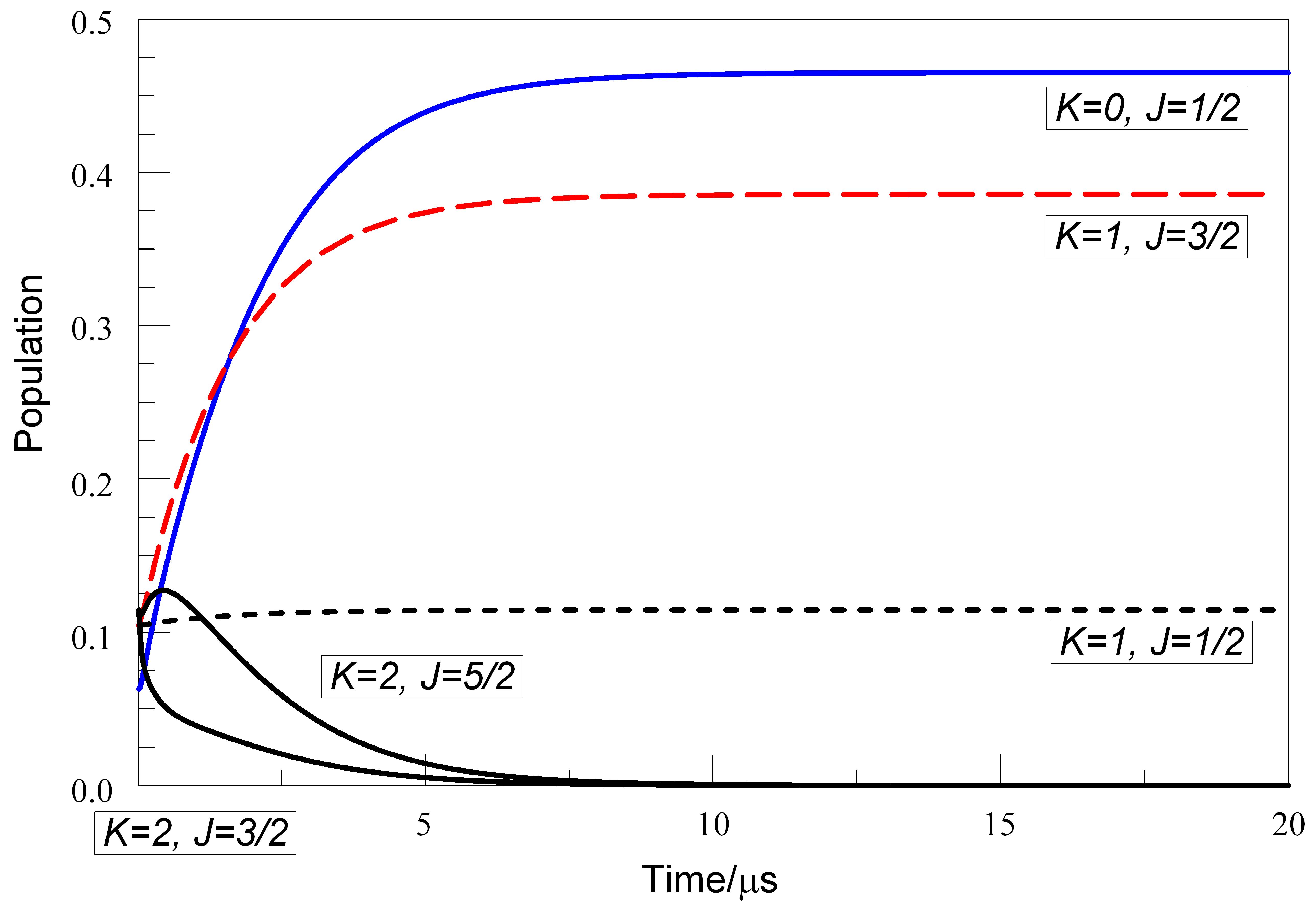}}
}
\mbox{
\subfigure[]{\includegraphics[height= 5.4cm]{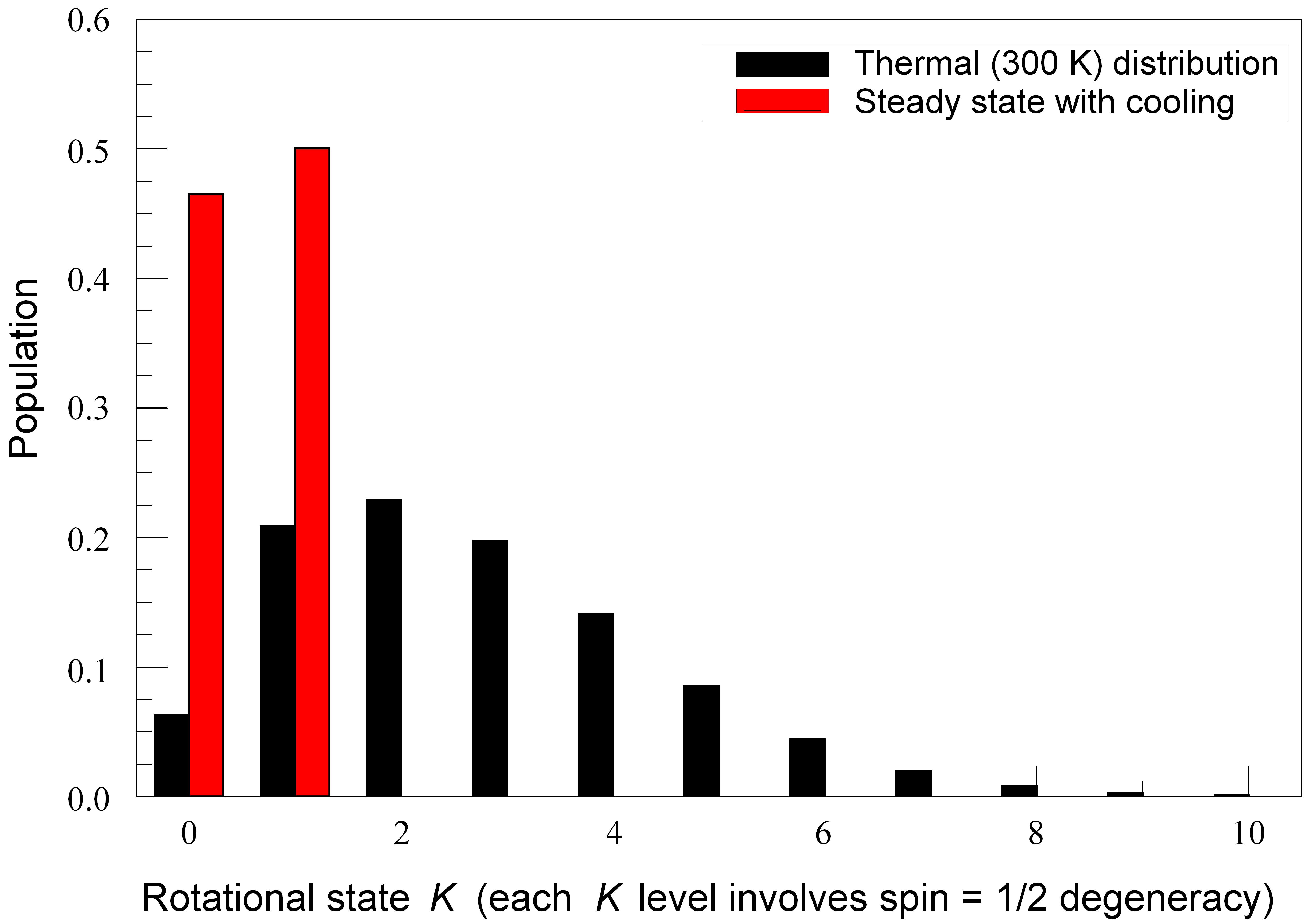}}
 }
\caption{Simulation of AlH$^{+}$ rotational cooling. (a) Simulation results of PROC for the time evolution of the lowest five spin-rotational levels. The population distribution reaches within 99$\%$ of equilibrium in 8 $\mu$s. (b) Population distribution among rotational states before and after PROC.}
\end{center}
\end{figure}

\section*{Rate equation simulation}

We modeled PROC for AlH$^{+}$ using a rate-equation simulation. Results are shown in Figure 3. Because of the highly diagonal FCFs of AlH$^{+}$ (Table 1), population leaks to higher vibrational states are negligible, and only one PFL is required. Vibrational population leaks are nonetheless included in the simulation. The simulation shows that PROC achieves 99$\%$ of population saturation in 8 $\mu$s, with an accumulated population in the lowest two states of 96.6$\%$. The different equilibrium populations of the dark states in Figure 3 reflect the number of spontaneous decay channels feeding into each state.

\begin{figure}[ht]                  
\centering
\includegraphics[height= 6cm]{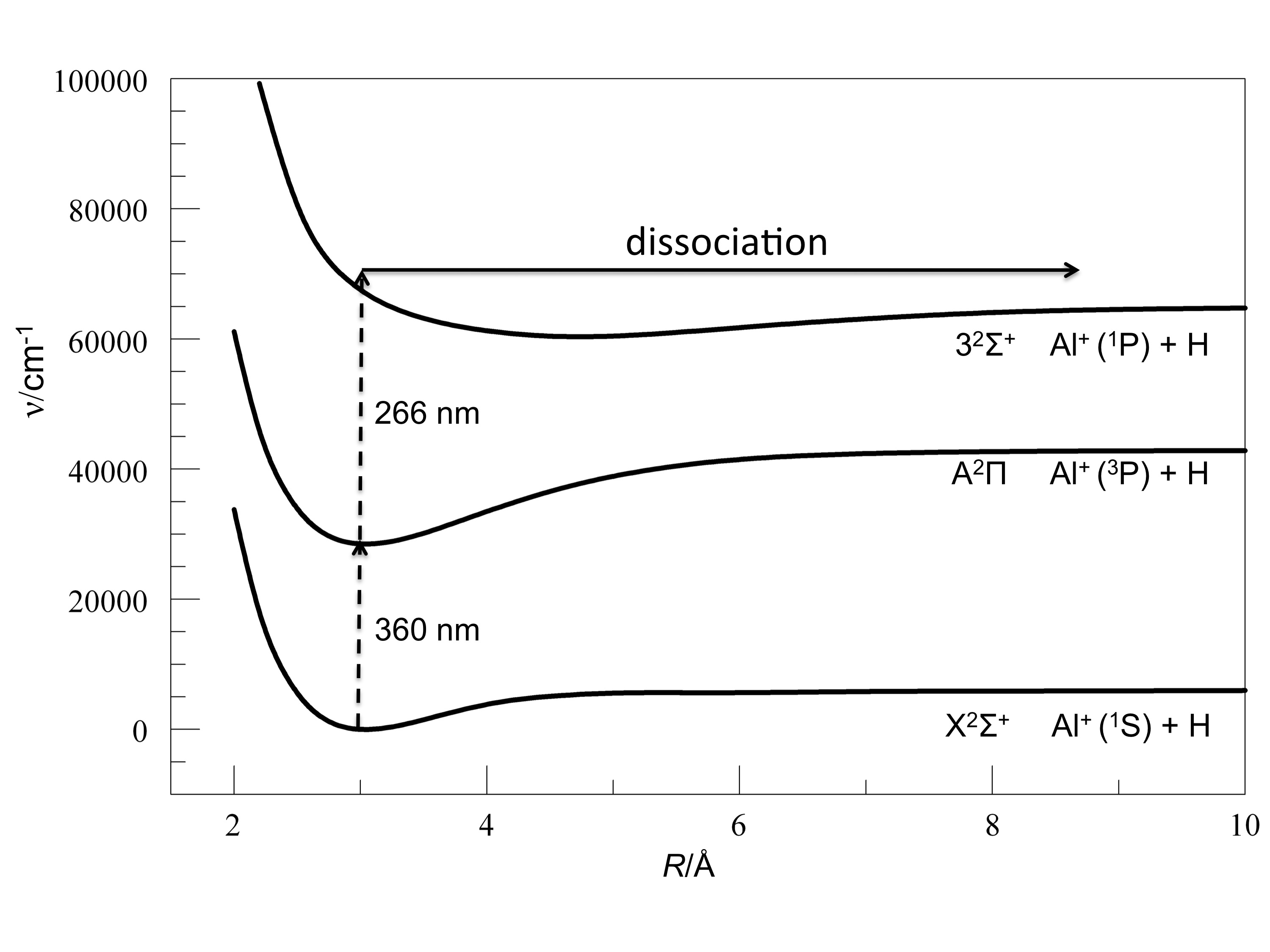}
\caption{AlH$^{+}$ rotational state readout scheme. The population in the X$^{2}\Sigma^{+} (v''=0)$ state is excited to a dissociating 3$^{2}\Sigma^{+}$ level, through the intermediate state A$^{2}\Pi (v'=0)$.}           
\end{figure}

\section*{State readout}

Evaluation of how well the molecular ions are prepared in the ground states is crucial; an accurate method is required to determine the population distribution among the rovibrational states. To this end, various approaches involving state-selective resonance-enhanced multi-photon dissociation (REMPD)\cite{Staanum2010, Schneider2010} and chemical reactions\cite{Tong2010} have been used. We intend to use an already demonstrated method for state readout that counts the number of state-selectively photodissociated ions by sequentially dumping different species from the trap onto a detector\cite{Removille2009}. AlH$^{+}$ ions in a specific rotational state of X$^{2}\Sigma^{+} (v''=0)$ will be optically excited to A$^{2}\Pi (v'=0)$ with a tunable pulsed dye laser at a wavelength of 360 nm, and then excited to a dissociating state (3$^{2}\Sigma^{+}$) with a 266 nm pulse generated from the 4th-harmonic of a Nd:YAG laser. Because the first photon excites the population from one bound state to another, the excitation process is state-selective. The two-photon-excited AlH$^{+}$ molecular ions dissociate into Al$^{+}$ ions and H atoms\cite{Guest1981, Almy1934, Bruna2003}. The Al$^{+}$ ions can then be extracted and counted, by applying a resonant radial secular excitation, which causes Al$^+$ to be transferred to an ion guide leading to a micro-channel plate (MCP). The secular excitation is mass-selective and can extract only the daughter ions without extracting the non-dissociated parent ions\cite{Removille2009}.

\section*{Conclusions}
In this article, we have proposed a practical scheme for realizing internal cooling of certain molecules with diagonal FCFs. The proposed PROC scheme is expected to achieve rovibrational cooling typically in several microseconds, without waiting for BBR population redistribution. A rate-equation simulation shows that PROC can cool AlH$^+$ to the two lowest spin-rotational states in 8 $\mu$s. The femtosecond laser used in our lab has a tunable range from 690 - 1040 nm, and can generate a wide range of broadband light by frequency multiplying ($e.g.$, SHG) or spontaneous parametric down-conversion. PROC is not applicable to all molecules; diagonal FCFs are required so that the primary PFL does not readily populate $v''>0$ states. Extra PFLs could be added for vibrational repumping, but nearly diagonal FCFs are still a practical requirement. PROC is applicable to polar and apolar molecules, potentially with reduced mass up to a few tens of Da, depending on the pulse-shaping resolution achieved. Currently we have demonstrated 360 nm pulse-shaping of a frequency-doubled femtosecond laser with sufficient bandwidth, power, and pulse-shaping resolution for rotational cooling AlH$^+$, BH$^+$, or SiO$^+$.
\\
\\

This work is supported by AFOSR (Grant No. FA9550-10-1-0221) and NSF CAREER (Grant No. PHY08-47748).
\\

\bibliography{PROC}
\end{document}